\documentclass[sigconf]{acmart}
\AtBeginDocument{%
  }

\setcopyright{acmlicensed}

\copyrightyear{2025} 
\acmYear{2025} 
\setcopyright{licensedusgovmixed}\acmConference[CCS '25]{Proceedings of the 2025 ACM SIGSAC Conference on Computer and Communications Security}{October 13--17, 2025}{Taipei, Taiwan}
\acmBooktitle{Proceedings of the 2025 ACM SIGSAC Conference on Computer and Communications Security (CCS '25), October 13--17, 2025, Taipei, Taiwan}
\acmDOI{10.1145/3719027.3765065}
\acmISBN{979-8-4007-1525-9/2025/10}

\usepackage{array}
\usepackage[table]{xcolor}
\usepackage{tabularx}
\usepackage{ragged2e}
\usepackage[style=american,threshold=2]{csquotes}
\newcommand\lorenzo[1]{{{\color{black}#1}}}

\begin{document}

\title{Extended Version: It Should Be Easy but... New Users' Experiences and Challenges with Secret Management Tools}

\thanks{\textcolor{red}{\textbf{A shortened version of this paper appears in the Proceedings of the 2025 ACM SIGSAC Conference on Computer and Communications Security (ACM CCS 2025), Taipei, Taiwan. This is the full version with the interview guide, participants demographics, and codebook in the appendix. All additional study materials are available at {\url{https://doi.org/10.5281/zenodo.17036161}}.}}}

\author{Lorenzo Neil}
\affiliation{
  \institution{North Carolina State University}
  \city{Raleigh}
  \country{United States of America}
}
\email{lcneil@ncsu.edu}

\author{Deepthi Mungara}
\affiliation{
  \institution{Paderborn University}
  \city{Paderborn}
  \country{Germany}
}
\email{deepthi.mungara@uni-paderborn.de}

\author{Laurie Williams}
\affiliation{
  \institution{North Carolina State University}
  \city{Raleigh}
  \country{United States of America}
}
\email{lawilli3@ncsu.edu}

\author{Yasemin Acar}
\affiliation{
  \institution{Paderborn University \& The George Washington University}
  \city{Paderborn}
  \country{Germany}
}
\email{yasemin.acar@uni-paderborn.de}

\author{Bradley Reaves}
\affiliation{
  \institution{North Carolina State University}
  \city{Raleigh}
  \country{United States of America}
}
\email{bgreaves@ncsu.edu}

\renewcommand{\shortauthors}{Lorenzo Neil et al.}

\begin{abstract}
Software developers face risks of leaking their software secrets, such as API keys or passwords, which can result in significant harm. Secret management tools (SMTs), such as HashiCorp Vault Secrets or Infisical, are highly recommended by industry, academia, and security guidelines to manage secrets securely. 
SMTs are designed to help developers secure their secrets in a central location, yet secrets leaks are still commonplace, and developers report difficulty in learning how to setup and use SMTs. While SMTs typically come with publicly available help resources (e.g., tool documentation and interfaces), it is unclear if these actually help developers learn to effectively use SMTs. Without usable help resources that onboards developers, quick adoption and effective use of SMTs may be unrealistic.

In a qualitative two-step study, we observed 21 new users in person while they used SMTs to perform two secret management tasks: secret storage and access, then secret injection. We interviewed participants after each task to identify their challenges and experiences using SMTs, with the assistance of help resources.

While our study sample is narrow, it serves as a reasonable proxy for new developers who are likely to adopt SMTs early in their careers.
We found that even in a laboratory setting where new users found tool functionality and interface flexibility helpful, they still experienced increased difficulty to effectively use SMTs to securely remediate a hard-coded secret when they felt tool documentation was insufficient. Insufficient tool documentation motivated participants to deviate from official tool documentation to access secondary sources or attempt workaround methods. 
Specific challenges reported by participants were tool documentation content quality, navigation difficulties with both tool documentation and web interfaces for finding helpful content, and supportive tool features. We explain how these challenges negatively affect participant experiences adopting SMTs, and suggest recommendations on tool documentation and interfaces for SMT developers. If developers cannot simply and quickly manage secrets securely, secret leakage will continue to be commonplace.
\end{abstract}




\maketitle
\section{Introduction}
Software developers need to store software secrets (e.g., passwords, API keys), which grant access to sensitive or private information (e.g., account access, company bank information). Without a proper secret management approach in place, developers may hard-code secrets into source code, public repositories, or configuration files. A highly recommended option by industry, latest guidelines, and academic research~\cite{basak2022practices, basak2023challenges, krause2023pushed} are secret management tools (SMTs). These can securely store sensitive developer secrets in a centralized location. However, even with the existence of SMTs, developers still leak secrets~\cite{SecretSprawl,basak2023challenges,krause2023pushed,dahlmanns2023secrets,AWSUrge}, which can lead to a wide exposure of sensitive information~\cite{Toyota,NyTimesGithubToken,meli2019bad,Demirkapi,DockerImages}.

Like most software, SMTs come with online public documentation that onboards developers on how to store, manage, and distribute secrets. Prior research efforts have highlighted issues within software tool documentation that affect developers' abilities to use the tools to perform software development related tasks~\cite{parnin2012crowd,nassif2023field,robillard2009makes,ernst2023study, acar2016you, acar2017comparing, fourne2024these}. Limitations for software development tool documentation may extend to SMT tool documentation, yet the extent of the impact of the quality of SMT tool documentation on developers' experiences in using SMTs is unclear. Prior research shows that SMT adoption is widely a burden for developers, largely because of setup and learning constraints during initial adoption of the tools~\cite{basak2023challenges,krause2023pushed}. If SMT tool documentation cannot minimize onboarding constraints, then new users will struggle to effectively use SMTs to centrally secure their secrets.

To understand the experiences and challenges new users face when using SMTs, we ran a qualitative two-step laboratory study to investigate the following research questions:

\noindent \textbf{RQ1:} \textit{Can new users effectively use a SMT to securely remediate a hard-coded secret?}

\noindent \textbf{RQ2:} \textit{What factors encourage or inhibit SMT usage success?}

We observed participants with prior secret management experience, using one of three cross-cloud SMTs: Doppler~\cite{doppler}, HashiCorp Vault Secrets~\cite{hcpsecrets}, and Infisical~\cite{infisical}. Participants performed two tasks that involved securing a hard-coded secret: Access a stored secret from the command line, then inject the secret into a local Python application. We chose these tasks to reflect a realistic onboarding scenario where a developer addresses a common issue involving a hard-coded database password in the codebase. This creates security risks through version control or logs~\cite{mykhaylova2024hardcoded,lykousas2024decoding}. The study tasks involve securing the secret using an SMT and injecting it at runtime, covering essential skills such as secret storage, access, and cloud integration. These tasks are based on real workflows and demonstrate how developers use secret management tools to reduce vulnerabilities~\cite{owasp_secrets_management_cheat_sheet}. Following each task, we interviewed participants about their experiences and challenges using SMTs and SMT help resources.

Although documentation had always been a consideration in our study design, our focus on it sharpened during the study based on participant responses (Section~\ref{sec:ToolDocSpecificChallenges}). We observed that participants were generally satisfied with the different functionality and user interfaces provided by the SMTs. While we noticed observable differences in task completion times between SMTs, all were functionally indistinguishable towards task success. However, even while appreciating SMT functionalities and interfaces, new users in our study struggled to effectively use SMTs to securely remediate a hard-coded secret. Participant interviews revealed that their ability to fully remediate the hard-coded secret was mostly affected by the quality of the available tool documentation.

Major limitations in SMT tool documentation that negatively impacted SMT usage were the following: Insufficient information for relevant command line arguments, lack of a relevant set of use cases in examples provided for command line interface (CLI) commands, and inconsistent versions of descriptions for CLI commands maintained across documentation. Participants also reported a lack of clarity when using the tool documentation and web interfaces to find technical guidance such as specific commands or debug information. When encountering limiting factors in SMT tool documentation, participants often deviated from tool documentation to access secondary sources and attempt workaround methods. However, participants largely reported that secondary sources were less helpful for their specific needs and therefore experienced more time and effort to troubleshoot and explore solutions. Our study conclusions contribute towards addressing research areas both specific towards SMTs~\cite{krause2023pushed,basak2022practices,basak2023challenges} and more broadly such as tool usability~\cite{8904489,acar2017comparing,kaiser2002evaluating} and documentation~\cite{10.5555/827987.827991,9470906,6958422}. Our work can serve as a reference for SMT vendors who need to create effective tool documentation and more user-friendly tool interfaces for their intended audience, regardless of their technical expertise, to enable broad, secure adoption of effective secret management tools. Thus, contributing to ensure secret leakage becomes less widespread.

\section{Background and Related Work}
We present a brief background on SMTs and discuss related work for secret management, software tool documentation, and security tool usability and adoption.

\subsection{Secret Management}
In response to secret leakage, recommended guidelines for software secret management from organizations~\cite{NISTSupplyChain,ReverseLabs}, coding platforms~\cite{gitGuardSecretMan,githubActions,githubCodeSpace}, and security blogs~\cite{SecretSprawl} were created to guide developers on how to secure secrets. Academic research has also addressed secret management leakage and practices. Meli et al.~\cite{meli2019bad} conducted studies to characterize widespread secret leakage in public repositories. Similar work has outlined solutions~\cite{sinha2015detecting,basak2022practices,tahaei2019survey} and proposed datasets~\cite{basak2023secretbench} to help developers detect and prevent exposed secrets in their repositories.
Basak et al.~\cite{basak2023challenges} characterized challenges and solutions reported in questions related to checked-in secrets on Stack Exchange, while Krause et al.~\cite{krause2023pushed} surveyed 109 developers and interviewed 14 developers to learn common challenges with managing and leaking secrets in code repositories. 

Both of their works motivate this study, as they highlight that developers want to use SMTs but struggle with adoption, often citing a lack of available documentation as a key challenge~\cite{basak2023challenges,krause2023pushed}. We further this effort by observing how participants use SMTs and the challenges they face in effectively managing secrets in direct relation to the available documentation.

\subsection{Secret Management Tools}
Secret management tools are highly recommended for securely storing sensitive software secrets in a centralized location~\cite{bestTools,G2}. SMTs can offer different functionalities or use cases. Cloud-specific SMTs are built on major cloud ecosystems such as AWS~\cite{aws-kms}, Google Cloud Platform~\cite{GoogleKMS}, or Microsoft Azure~\cite{azure}, and are typically used by companies already integrated into those ecosystems. On the other hand, Cross-cloud SMTs are usually offered either as Software-as-a-Service (SaaS) or as a managed service that can be self-hosted or managed~\cite{bestTools}. Cross-cloud SMTs are not tied to a specific cloud and offer easier onboarding for developers not working within large cloud ecosystems. Open source SMTs are self-hosted by developers, while in-house SMTs are custom-built by companies for their own use. While both give developers much creative freedom, these services must be regularly maintained by the company or developers themselves, therefore requiring significant overhead. Interview findings from Krause et al.~\cite{krause2023pushed} show that developers report setup and learning constraints when first adopting SMTs. For our work, we investigate new user experiences with three cross-cloud SMTs and the respective tool documentation.

\subsection{Software Tool Documentation}
Much of the prior work towards software documentation specifically investigated challenges with API focused software documentation that prevents developers from effectively and securely using APIs~\cite{uddin2015api,acar2017developers,sohan2017study,treude2016augmenting,robillard2015recommending,head2018not,acar2016you,subramanian2014live, acar2017comparing}. Challenges highlighted from prior work that affect the quality of software documentation include, but are not limited to, insufficient coding examples~\cite{aghajani2020software,aghajani2019software,nassif2022casdoc,acar2017developers,acar2017comparing,subramanian2014live}, incorrect or ambiguous explanations~\cite{uddin2015api,wen2019large,middleton2020data,chen2009empirical}, as well as non intuitive presentation of material~\cite{nassif2023non,nassif2023field}. Such challenges within software documentation make it more difficult for developers to learn new technology~\cite{nassif2023identifying,arya2023programmers,chen2009empirical} and in some cases affect developer productivity~\cite{wagner2019factors,segal2007some,murphy2019predicts,head2018not,noda2023devex}. In 2003, Lethbridge et al.~\cite{lethbridge2003software} interviewed software engineers to learn how they used software documentation. 61\% of the participants in their study felt software documentation was most effective when learning new software and that they prefer simplified documentation, while tending to ignore complex or time consuming documentation~\cite{lethbridge2003software}. More recent work highlights how developers seek both, official documentation and also other resources such as Stack Overflow~\cite{arya2023programmers,baltes2020contextual,robinson2022error,storey2024disruptive,acar2016you} to learn new technology. Stack Overflow in particular has been highlighted in prior work as a resource developers access to find answers to challenges addressed by their community~\cite{storey2024disruptive,parnin2012crowd}. Parnin et al.~\cite{parnin2012crowd} examined crowd contributed API documentation from Stack Overflow and found that while the crowd documentation provided many examples and explanations for API elements, the rate of information produced from the crowd would not be fast enough to outright replace API documentation~\cite{parnin2012crowd}. A lab study conducted by Acar et al.~\cite{acar2016you} observed Android developers writing security-and-privacy-relevant code under time constraints using different information sources. Their findings reported that developers who used Stack Overflow as a resource wrote less secure but more functional code as opposed to developers who used official Android documentation~\cite{acar2016you}. While we see the critical need to create effective software documentation, there is a lack of incentive or motivation to create such documentation~\cite{parnin2012crowd,arya2024people}. Arya et al.~\cite{arya2024people} interviewed 26 volunteer documentation contributors to learn their motivations for contributing to software documentation. Their work concluded that contributors to software documentation were mostly self-motivated, including personal experiences with inadequate software documentation or pursuits for content creation. 

Challenges with software documentation content, presentation, and availability as highlighted from prior work serve as motivation for this work. We extend the body of literature by observing how challenges for SMT tool documentation affect our participants' abilities to use SMTs to perform tasks involving managing secrets. We also observe how tool documentation content challenges motivate participants to find secondary sources for further assistance or try workaround methods.

\subsection{Security Tool Usability and Adoption}
The usability of security tools impacts developers' ability to securely complete tasks~\cite{storey2019towards,cheng2022improves,acar2017comparing,witschey2015quantifying}. Storey et al.~\cite{storey2019towards} surveyed software developers at a company to learn about challenges they faced in regards to their job satisfaction and perceived productivity. Challenges with software architecture and finding relevant information were two of the most common challenges their survey participants reported in regards to their job satisfaction and perceived productivity. Acar et al.~\cite{acar2017comparing} compared the usage of different cryptographic API usage by Python developers to write secure code and found that simplified libraries within the APIs helped developers produce more secure code than the comprehensive libraries. Related prior work also echoes the theme that simplifying information or the design of tools improves tool usability~\cite{smith2020can,indela2016helping}. In response, prior work has suggested new approaches aimed to help developers write more secure code when using cryptographic APIs~\cite{gorski2018developers,kruger2017cognicrypt}. Gorski et al.~\cite{gorski2018developers} observed improvements in code security when comparing developers who used Python's PyCrypto API to developers with a version of PyCrypto with integrated security advice. Krüger et al.~\cite{kruger2017cognicrypt, kruger2023securing} introduced CogniCrypt, their proposed tool for helping developers write code and implement tasks in a secure manner. 

Much of the prior efforts on examining security tool usability and adoption focused on APIs or other specific areas. We use prior work as an influence to shape how we investigate the usability of SMTs, which has only been briefly discussed in prior work~\cite{krause2023pushed,basak2023challenges}. We also examine how provided tool documentation and the availability of secondary sources impact the ability of participants to use SMTs for managing secrets.

\section{Methodology}
We conducted a qualitative two-step study on SMT use with 21 Computer Science (CS) Master’s students in July 2024, consisting of an observation study followed by an interview study~\cite{article, Jamshed2014}. Participants were observed while using SMTs, with the assistance of publicly available tool documentation from the SMTs, to perform two tasks: Task 1 involves storing and accessing a secret and Task 2 involves injecting a secret into a local application. After each task, we interviewed participants to learn about their experiences and any challenges they faced while completing each task. In this section, we explain the methodology of our study. 

\subsection{Recruitment and Eligibility Criteria}
\label{sec:ParticipantRecruitment}
Our population of interest were developers with academic or professional experience with managing secrets for coding projects. For feasibility, we focus our recruitment towards a sub-population: CS Master's students at our university. Many local CS Master's students either have experience managing secrets from their coding projects, prior employment positions, research projects, or academic courses, and are therefore an appropriate sample for our study. We focused on participants with prior secret management experience, so that true new SMT users can be expected to fare either the same or worse.

We recruited CS Master's students through a public email forum offered by our CS graduate office. We described our research overview and goals, and provided a link to our Qualtrics~\cite{qualtrics} screening survey for participants to indicate interest. We required participants to participate in person at our research campus and have either prior experience or general understanding of the following: Linux or MacOS terminal commands, Python file execution, environment variable usage, and secret management, all essential for performing our research tasks. Participants were only eligible if they met all of our criteria. Once eligible participants filled out our screening survey, we sent them a link to our Qualtrics~\cite{qualtrics} informed consent form. Our informed consent form asked for consent to participate in person and agree to be recorded through both video and audio. Participants were compensated \$30 per half hour for their participation in the study. 

We ended participant recruitment when we observed common usability issues and themes were repeated, after observing no new insights, from participants’ performances to address \textbf{RQ1}, and from participants’ interview responses to address \textbf{RQ2}.
Participant demographics in Appendix~\ref{demographics} shows the breakdown of their prior professional experience, academic experience, gender, and the SMT they were assigned to use in the study. All but one of our participants had prior experience with managing secrets in a company, either through a prior internship or full time employment.
Nine participants held a combination of both professional experience and academic coursework experience with managing secrets. They named passwords, tokens, API keys, and user credentials as examples of secrets they were tasked with managing in prior experiences. Participants mentioned using the secret management tools Amazon Web Services (AWS) Key Management Service~\cite{aws-kms}, Azure Key Vault~\cite{azure}, and Google Cloud Platform (GCP)~\cite{GoogleKMS}.

\subsection{Study Design}
\label{sec:StudyDesign}

\subsubsection{SMT Choice} 
\label{sec:SMTTools}
We chose the following three cross-cloud SMTs:  Doppler~\cite{doppler}, HashiCorp Vault (HCP) Vault-Secrets~\cite{hcpsecrets}, and Infisical~\cite{infisical}. We chose cross-cloud SMTs because they do not require substantial experience or access with specific cloud providers. Hence, we avoided cloud-specific tools, such as AWS, which require substantial experience with that particular cloud deployment.

All three SMTs offer public online tool documentation and free versions of their tools. 
These three specific cross-cloud SMTs were also three of the more common cross-cloud SMTs~\cite{bestTools, G2}. All three SMTs provide a web user-interface (UI) and installable command-line interface (CLI). The web UI allows users to create free accounts and store secrets. Secret storage dashboards allow users to create and manage projects that store their secrets. From the CLI perspective, users can access secrets from their web secret storage projects and distribute them across local applications without the need to store secrets locally.

\subsubsection{Study Protocol}
\label{sec:StudyProtocol}
Our study uses a between-subjects research design~\cite{betweenvwithin}, where each participant met with a researcher on campus and performed two research tasks with only one of the three SMTs. Therefore, seven participants performed two secret management related tasks with each SMT, totaling 21 participants overall. We present a table that lists which SMTs participants used for our study in Table~\ref{fig:participantExp} and the study protocol details can be found in our replication package\footnote{\url{https://doi.org/10.5281/zenodo.17036161}}.

\begin{figure}
\begin{center}
\includegraphics[width=\linewidth]{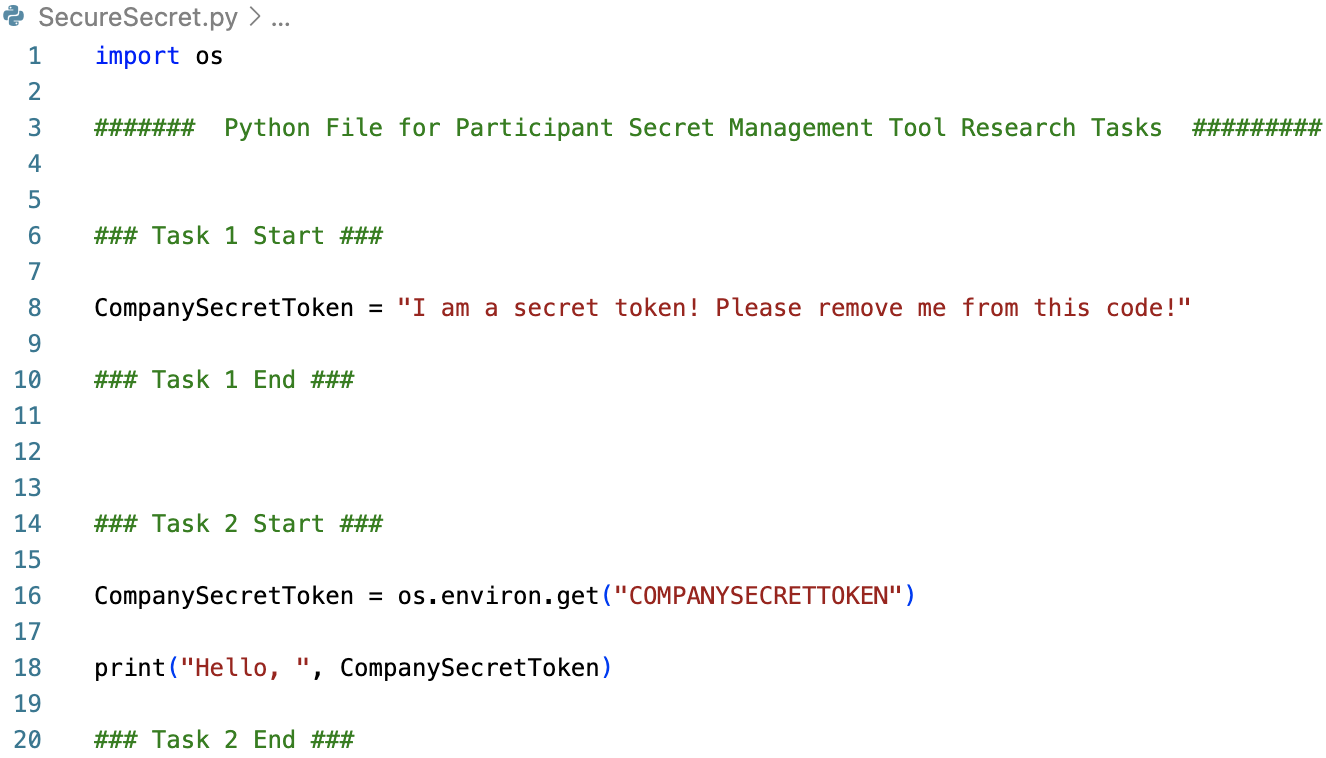}
\caption{Screenshot image of the Python file, \texttt{SecureSecret.py}, used by participants for Task 1 and 2.}
\label{fig:PythonFile}
\end{center}
\end{figure}

We borrow inspiration from related work by providing participants with a scenario-based description for each task~\cite{acar2016you,oh2024poisoned,bai2016inconvenient}. Each scenario imagines the participant working for a software company with the goal to use the provided SMT, with respective tool documentation, to remediate a hard-coded secret within a company Python file shown in Figure~\ref{fig:PythonFile}. The Python file, \texttt{SecureSecret.py}, consists of one or two lines of code for each task. One of the lines of code for Task 1 includes the hard-coded secret \texttt{CompanySecretToken}. Participants were provided with a MacBook Pro which held the Python file in a Visual Studio Code (VS) Text Editor work space~\cite{vsCode}. Participants also were required to use the terminal prompt within VS to enter SMT-based CLI prompts to complete steps for each task. Lastly, participants were given account access to the web UI for their SMT which held the scenario-based project, ``\textit{Company Secret Storage}'', they store \texttt{CompanySecretToken} in. 

While we described the first task to participants, we provided them links to online tool documentation from the SMT they worked with. The links explicitly provide information for each task within their core CLI commands documentation, quick starter guides, and documentation that covers managing secrets or projects. We present these links to the participant in a Google Doc where we also provide labels for the titles of each link. The list of links we presented for Doppler, HCP Vault Secrets, and Infisical can be found in our replication package.  
We did this to help the participants reduce trivial lookup time and decrease the disadvantages of finding advice on different SMTs. 

Participants were given 30 minutes to complete each task, a duration chosen based on pilot testing and complexity of tasks. We allowed stoppages of time when participants asked questions or wanted to take a break. We also allowed participants to skip a task or not answer a question. If participants reached a point in which they felt the SMT tool documentation was unclear or not helpful, we allowed participants to search online for secondary sources for more information. We allowed participants to use any website (e.g., YouTube or Stack Overflow) or medium (e.g., text or video). However, we did not allow participants to use generative AI services (e.g., ChatGPT). We set this restriction since not all participants may be familiar with such services and we wanted to ensure participants have the same resource availability and all know how to use their resources. 

After participants completed a task, we asked them several questions about their experiences and challenges using the SMT and tool documentation. Participants were also asked a couple of warm-up questions before the tasks, which collected their prior experience with secret management. After participants answered questions about both tasks, they were then asked a set of wrap-up questions. We explain in detail our post task interview development and structure in Section~\ref{sec:InterviewDevelopment} and ~\ref{sec:InterviewStructure}.

\subsubsection{Task Selection} 
\label{sec:TaskSelection}
We designed each task to be general and be equally applicable to all SMTs we study and to reflect real-world security practices, focusing on secure secret management and minimizing the risk of hardcoded secrets, a common security vulnerability~\cite{mykhaylova2024hardcoded, baskaran2023measuring}. These tasks also reflect common, security-critical workflows that developers regularly perform when managing secrets in the development and deployment process~\cite{owasp_secrets_management_cheat_sheet}. Using both the UI and CLI as part of our task design enables us to uncover usability issues that could affect whether developers adopt security tools and how they use them over time. Both tasks were designed to be completed within 30 minutes. By evaluating how participants handle secret storage, retrieval, and injection, our study addresses the security and usability perceptions of developers when using SMTs. Participants also do not need extensive knowledge of the SMT to perform these tasks. With the tool documentation provided to participants, as well as secondary sources, participants were instructed to complete the following two tasks:

\textbf{Task 1:} Remove \texttt{CompanySecretToken} from \texttt{SecureSecret.py}, store \texttt{CompanySecretToken} within \textit{Company Secret Storage}, and then access \texttt{CompanySecretToken} from the command line using the SMT CLI. 

\textbf{Task 2:} Inject \texttt{CompanySecretToken} into \texttt{SecureSecret.py}.

\begin{figure}
\begin{center}
\includegraphics[width=\linewidth]{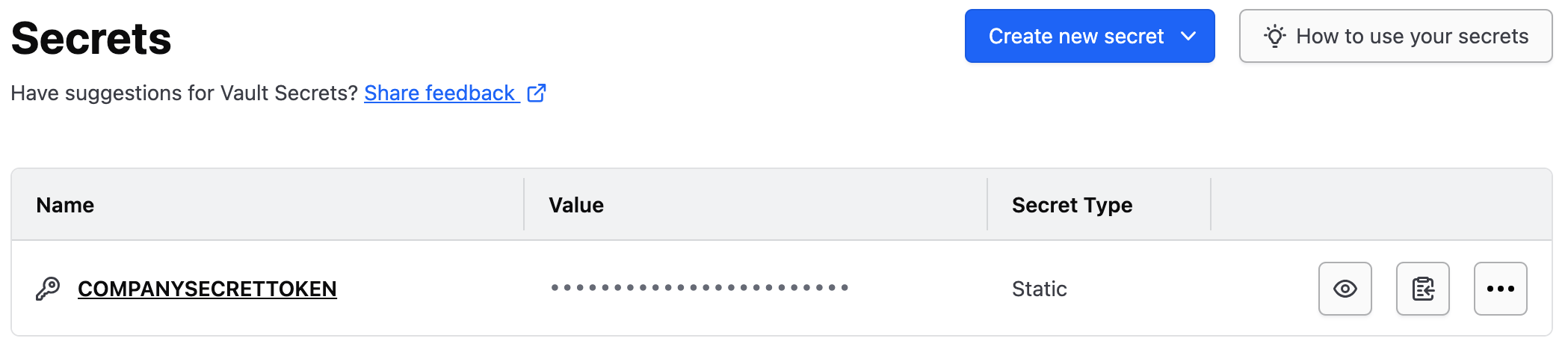}
\caption{Screenshot image showing the HCP Vault Secrets dashboard for the \textit{Company Secret Storage} project with \texttt{CompanySecretToken} stored.}
\label{fig:hcpCompanySecretStorage}
\end{center}
\end{figure}

\begin{figure}
\begin{center}
\includegraphics[width=\linewidth]{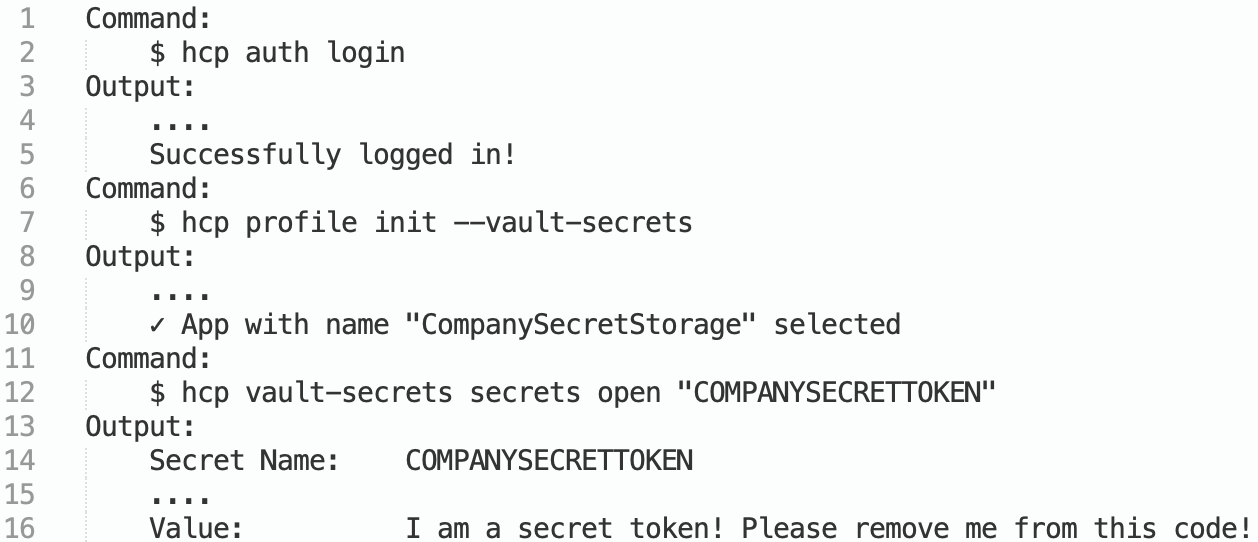}
\caption{Sample presentation of the list of CLI commands required to complete Task 1 for HCP Vault Secrets.}
\label{fig:hcpCLITask1}
\end{center}
\end{figure}

Task 1 starts with participants removing \texttt{CompanySecretToken} from \texttt{SecureSecret.py}, as shown in Figure~\ref{fig:PythonFile}. For this, we simply have them delete it from the file. Participants then store \\ \texttt{CompanySecretToken} using web UI of a SMT in the project named \textit{Company Secret Storage} as shown in Figure~\ref{fig:hcpCompanySecretStorage}. The last step for Task 1 is to then use the SMT CLI to print the plaintext value of \texttt{CompanySecretToken}. A presentation for the number of steps required for this task from the CLI perspective is shown in Figure~\ref{fig:hcpCLITask1}.

Task 2 involves a process called ``secret injection''~\cite{hcpVaultSecretsRun}. For secret injection, a secret is retrieved from the SMT's central storage, and then passed to an application at run time as an environment variable. Secret injection ensures that the secret is not stored locally or within the application. As shown in Figure~\ref{fig:PythonFile}, there are two lines under Task 2 within \texttt{SecureSecret.py}. The first line retrieves an environment variable named \texttt{CompanySecretToken}. Once the participant completes Task 1, they would have successfully logged in and initialized the SMT with the CLI. Participants can then reference secrets stored in the SMT from the CLI through the usage of environment variables. We take care of this step with the first line of code which creates the environment variable. The second line simply prints ``\texttt{Hello, }'' in front of the value of the environment variable we created for \texttt{CompanySecretToken}. 

\begin{figure}
\begin{center}
\includegraphics[width=\linewidth]{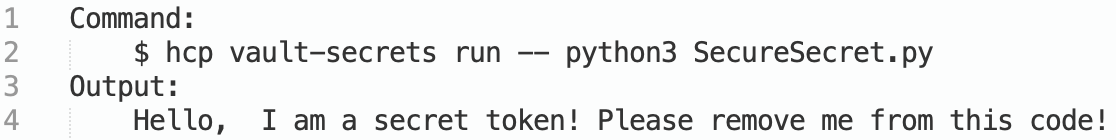}
\caption{Sample presentation of the injection run CLI command required to complete Task 2 for HCP Vault Secrets.}
\label{fig:hcpCLITask2}
\end{center}
\end{figure}

After participants complete Task 1, injecting \\ \texttt{CompanySecretToken} only requires one command as shown in Figure~\ref{fig:hcpCLITask2}. This is common for all of the SMT's for each Task 2 as each SMT's injection command consists of a similar syntax which all commonly use the term ``run''. The ``run'' term is followed by two dashes and then the normal start command one would input to run the application in which they are injecting a secret into (e.g., \texttt{python3 SecureSecret.py} in our case). The expected output for Task 2 is: \texttt{Hello, I am a secret token! Please remove me from this code!}.

\begin{figure}
\begin{center}
\includegraphics[width=\linewidth]{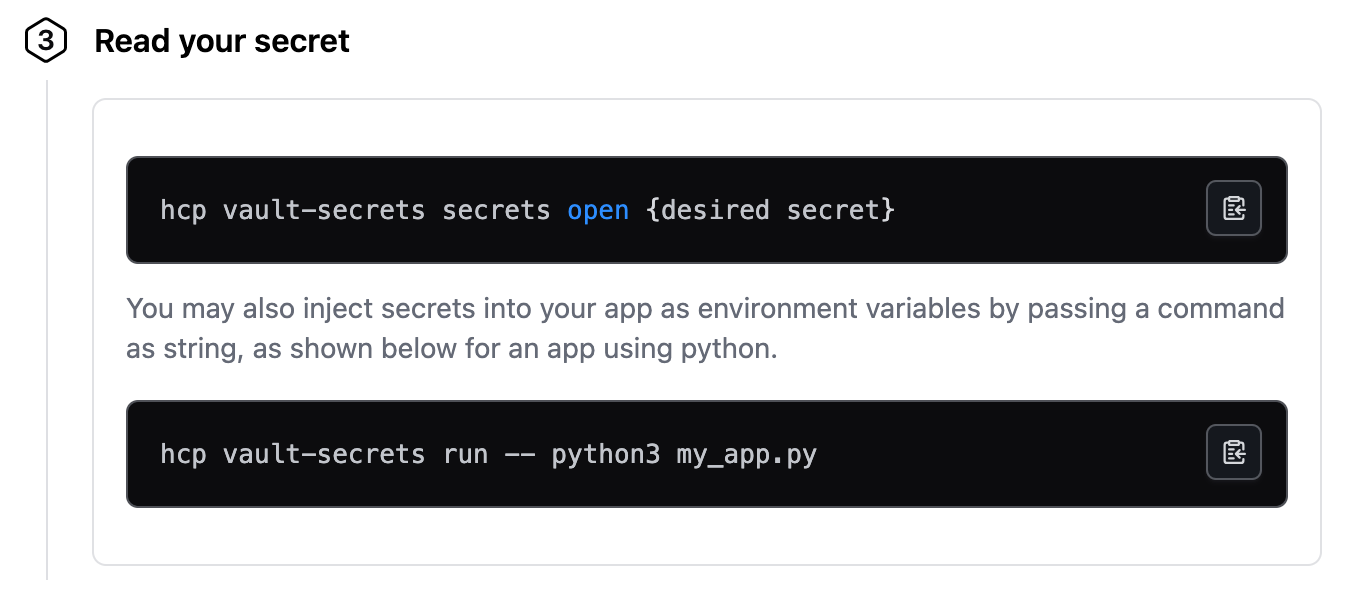}
\caption{Additional part of HashiCorp vault secrets documentation in their web dashboard that reflects the required CLI command for Task 2.}
\label{fig:hcpWebPortalPython}
\end{center}
\end{figure}

\textbf{Study considerations:} We initially chose a Python local file for Task 2 since before our pilot study, none of the SMTs held a specific Python example for their secret injection run advice. We also chose Python since it is the most popular coding language as well~\cite{IEEEPython,TIOBEPython,StackOverflowCodingPython}, and would provide a common language that developers code in already. We carefully designed these tasks to minimize any advantages between the SMTs and their tool documentation. However, after our pilot study, HCP Vault Secrets updated their documentation in their web dashboard to add an example of how to complete Task 2 in Python, as shown in Figure~\ref{fig:hcpWebPortalPython}. We also noticed that Doppler references an external PyPI webpage~\cite{DopplerPython} in their ``Development/Editors'' documentation which also shows an example of how to complete Task 2 in Python. Infisical throughout the time of our full study did not include such content in neither their public help documentation nor web UI. We decided to continue our experiments as planned and take note if these additions to the Doppler help documentation and HCP Vault Secrets web dashboard had any affect on the participants completing the tasks. We viewed this phenomenon as a natural experiment. Meaning, we observed differences in how participants performed on tasks using tool documentation when naturally presented with different content specificity.

\subsubsection{Data Collection}
\label{sec:StatsAnalysisMethods}
We recorded whether participants completed each task and their completion times for each task. We also recorded two specific actions from participants that indicate deviations from the expected or ``recommended'' path a user would take to complete these tasks in a real setting. The expected path we refer to is that participants complete a task using the SMT and steps recommended by the SMT tool documentation to complete each task within the time constraint. The two actions we recorded which represent a deviation from the expected path are usage of secondary sources and attempts at workaround methods.

The usage of secondary sources means that for a given task, a participant searched for and accessed a source outside of the tool documentation provided by the SMT in order to complete a task. By ``workaround methods'', we refer to trying commands or actions that were not the recommended method suggested by the SMT tool documentation to complete a task (e.g., exporting tokens to access secrets or using the injection command to print the plaintext version of a secret in the CLI in Task 1). These actions allowed us to ask follow-up questions in our post-task interviews to see why participants deviated from the expected path in order to complete a task. We present these findings in detail in Section~\ref{sec:QuantFindings}.

\subsection{Post-Task Interviews}

\subsubsection{Interview Guide Development}
\label{sec:InterviewDevelopment}
We interviewed participants after each task to get in-depth responses about their experiences during each task. We focused our questions on the usage of the SMT, tool documentation, and usage of secondary sources if they accessed them. To develop questions addressing this focus, we first referred to prior work on software documentation~\cite{parnin2012crowd,acar2017developers} and tool observation studies~\cite{robillard2009makes,acar2017comparing}. Prior work has highlighted content challenges such as insufficient code snippets, explanatory text, or lack of references for a wide range of use cases.  
Therefore, we used known related challenges from prior work as well as our research questions to draft several high-level categories of questions. We then piloted our interview guide with our observational study design with three PhD research students in our research lab to refine our interview guide and study design. Two of the PhD students pursue research in software engineering and software supply chain security, while the third PhD student pursues research in usable security and privacy. We updated our interview guide based on feedback from pilots. The complete interview guide is available in Appendix~\ref{sec:interview questions}.

\subsubsection{Interview Structure}
\label{sec:InterviewStructure}
Warm-up and wrap-up questions are asked before and after both tasks, respectively. The following high-level categories here represent questions we asked participants:

\textbf{Warm-up Questions:} Before Task 1, we asked a set of warm-up questions to learn about their academic and/or professional experience with secret management, as well as the types of secrets and tools they have used.

\textbf{Overall Experience:} The first question we asked after participants finished a task is to describe their overall experience with the task.

\textbf{Negative Emotions:} We asked participants if they felt any negative emotions when performing each task. Negative emotions can include annoyance, confusion, fatigue, frustration, or any form of disapproval/dissatisfaction. This focus helps identify usability issues that may not emerge from general feedback, while positive experiences were captured in the overall experience question.

\textbf{Documentation Content:} We specifically asked participants about challenges they experienced using the content of the tool documentation itself to complete each task. Content includes (but is not strictly limited to) code snippets, code blocks, CLI commands, coding or web UI examples, explanatory text, images, or videos. 

\textbf{Documentation Structure and Formatting:} We specifically asked participants about challenges they experienced with the structure and formatting of the tool documentation. Structure includes (but is not strictly limited to) section headers, bullet points, tabs, or toolbars. Formatting includes (but is not strictly limited to) text, pictures, or videos used in the advice.

\textbf{Secondary Source Usage:} We then asked participants about their usage of secondary sources and if those sources were more challenging to use or more helpful than the official tool documentation.

\textbf{Wrap-up Questions:} We wrap up the study by asking participants if they would consider using the SMT in the future, what their general preferences are for receiving information from sources, and for any last comments or questions about the study.

\subsubsection{Analysis and Coding}
\lorenzo{All interviews were recorded in person via Zoom and then transcribed using a Whisper’s multi-lingual medium model ~\cite{whisper} into written transcripts.} Interview transcripts were analyzed through qualitative deductive and inductive coding. We first developed an initial codebook based on our interview and research questions. Using this initial codebook, two of the researchers met to code a subset of the transcripts and compare their coding results. After the subset coding, the two researchers then split coding evenly (13 and 14 transcripts coded respectively) and met weekly to discuss coding disagreements and revise the codebook to accurately reflect our research interests. The two researchers also double coded six of the transcripts and checked inter-coder agreement weekly~\cite{mcdonald2019reliability} to ensure we accomplished high agreement when independently coding transcripts (Krippendorff’s Alpha > 0.75 for both researchers during independent coding~\cite{Krippendorf,Krippendorf3}). We continued to iteratively code the interview transcripts until no new codes emerged and all transcripts were coded. The final codebook can be found in Appendix~\ref{sec:InterviewCodebook}.

\subsection{Data Protection and Ethical Standards}
We took multiple steps to protect participants' privacy and data, while also ensuring all participants were treated ethically and with respect.
Before participants performed our study, we described our research goals to them and obtained consent from all participants. Participants were also informed on how their information would be protected on our consent form. During the study, we briefed participants on their rights during the study, including the right to skip any tasks and questions, and the right to withdraw without loss of benefits. We obtained video recordings and journal notes from each session, as well as written transcripts from audio recordings and performed both quantitative and qualitative analyses on our data. Participants consented to transcription, and all transcription was done with local models~\cite{whisper}. Written transcripts and journal notes were de-identified by replacing personally identifiable information (participant names, organizational names) with pseudonyms such as P01. We also did not request identifying, confidential, or private information about our participants or their prior experience in our study. We used end-to-end encrypted tools in all of our study communications and data storage components. Our study protocol was approved by our University's Institutional Review Board (IRB). 

Further, we ensured that none of the CS Master's students held any prior relationships (academic or professional) or interactions with the researcher leading the in-person observational tasks and interviews. We also informed participants they would only receive monetary compensation, and their involvement would not result in any academic credit compensation or hold any bearing on their academic status. 

\subsection{Limitations}
Participants were placed under three constraints during this study. The first being a 30 minute time constraint to complete each task. Such a time constraint may have influenced participants to overlook or skip steps that they normally would take in order to finish the task. Also, emotions of stress or fear of incompletion may also have been a factor for participants during this study. We informed participants that their compensation is not dependent on performance, but rather based on the amount of time they take to complete tasks. We also allowed participants breaks at any time, allowed questions in which we pause the time, or allowed them to skip a task if they chose to do so. The second constraint is that participants only used one of the three SMTs during their tasks. We designed our study this way because we did not want to overburden or confuse participants with too many tools or tasks, and also to avoid learning effects. Our third constraint prohibited using generative AI services when searching for online SMT tool documentation. Allowing AI tool use would introduce many  confounding variables. One issue is assessing and controlling for user experience and competence with AI. Another is accounting for tool and model choice, model performance, and if AIs are biased to perform better against older or more popular tools. Studying the effectiveness of GenAI for this setting will require novel methodologies worthy of their own future research efforts.

Our study sample consists of CS Master's students with mixed levels of secret management experience from professional or academic backgrounds, thus potentially contributing to confounding effects to our study by analyzing participants with varying levels of software development maturity. However, focusing on CS Master's students serves as a proxy for junior developers, which is reflective of the new user study population we focus on.

\section{Results}
This section presents the findings of the challenges and experiences participants faced when using an SMT to perform two secret management related tasks. In Section~\ref{sec:QuantFindings}, we report findings for participant task completion, completion times, usage of secondary sources, and attempts at workaround methods for each task. In Section~\ref{sec:TaskStatsCompare}, we present noticeable differences for participant performance for tasks across the three SMTs. The remaining sub-sections report in-depth qualitative findings from post-task interviews with participants who reported their challenges and experiences using SMTs to complete our secret management tasks. Our qualitative findings aim to explore the context for why developers widely report setup and learning constraints for SMTs when adopting them~\cite{krause2023pushed,basak2023challenges}. In Section~\ref{sec:WrapUp}, we report participants' insights and responses regarding the functional impact of the SMTs they used. In Section~\ref{sec:ToolDocSpecificChallenges}, we report specific challenges with the tool documentation that contributed to reported task challenges and deviations from the expected methods as provided from the tool documentation. Lastly, in Section~\ref{sec:SecSources}, we explain responses participants gave for why they accessed secondary sources, and also explain how helpful participants reported that access to secondary sources was for their tasks.

\subsection{Participant Performances}
\label{sec:QuantFindings}
Tables~\ref{fig:participantObsTask1} and ~\ref{fig:participantObsTask2} show the performance statistics for participants. Overall, we observe noticeable differences in participant performance between tasks. 21 participants completed Task 1 within 30 minutes, whereas only 18 participants completed Task 2 within 30 minutes. We also notice that participants on average completed Task 1 in 11 minutes and 8 seconds, whereas participants on average completed Task 2 in 16 minutes and 18 seconds. More importantly, we noticed that only eight participants accessed a secondary source or tried a workaround method in Task 1, while 16 participants accessed a secondary source or tried a workaround method in Task 2. In Task 2, 12 participants accessed a secondary source, and only two participants accessed a secondary sources in Task 1. Twice as many participants tried a workaround method in Task 2 (14) as they did in Task 1 (7). 
\begin{table}[ht]
    \centering
    \caption{Participants' Performance By SMT in Task 1.}
    \label{fig:participantObsTask1}
    
    \newcolumntype{R}{>{\RaggedLeft\arraybackslash}X}
    \newcolumntype{L}{>{\RaggedRight\arraybackslash}X}
    \rowcolors{2}{white}{gray!25}
    \begin{tabularx}{\columnwidth}{X R R R R R }
        \toprule
        \multicolumn{2}{r}{Completed} & SMT Used &  Secondary Source & W.A Method & Time\\
        \midrule 
        P16 &   $\checkmark$ &     HCP VS  &                &                &  1:52            \\
        P21 &   $\checkmark$ &     HCP VS  &                &                &  2:41             \\
        P12 &   $\checkmark$ &     HCP VS  &                &                &  3:34             \\
        P09 &   $\checkmark$ &     HCP VS  &                &                &  5:35             \\
        P03 &   $\checkmark$ &     HCP VS  &                &                &  7:10            \\
        P01 &   $\checkmark$ &     HCP VS  &                &                &  8:57            \\
        P11 &   $\checkmark$ &    HCP VS  &                &                &  14:03            \\
        P07 &   $\checkmark$ &     Infisical  &                &                &  5:40             \\
        P05 &   $\checkmark$ &     Infisical  &                &   $\checkmark$ &  7:56 \\
        P20 &   $\checkmark$ &     Infisical  &                &                &   8:31           \\
        P10 &   $\checkmark$ &    Infisical  &                &   $\checkmark$ &   10:39 \\
        P02 &   $\checkmark$ &    Infisical  &                &                &   17:03      \\
        P04 &   $\checkmark$ &    Infisical  &   $\checkmark$ &                &   19:45 \\
        P14 &   $\checkmark$ &    Infisical  &   $\checkmark$ &   $\checkmark$ &   23:42 \\
        P15 &   $\checkmark$ &     Doppler  &                &   $\checkmark$ &   6:44 \\
        P08 &   $\checkmark$ &     Doppler  &                &   $\checkmark$ &   9:43 \\
        P19 &   $\checkmark$ &    Doppler  &                &                &   10:33         \\
        P06 &   $\checkmark$ &    Doppler  &                &   $\checkmark$ &    11:52 \\
        P13 &   $\checkmark$ &    Doppler  &                &                &   12:17        \\
        P18 &   $\checkmark$ &    Doppler  &                &                &   22:07        \\
        P17 &   $\checkmark$ &    Doppler  &                &   $\checkmark$ &   23:28 \\
        \midrule
        Overall & 21  &   & 2      & 7      & 11:08$^\dagger$   \\ 
        \bottomrule
    \end{tabularx}
    \RaggedLeft
    $^\dagger$ Average
\end{table}

\begin{table}[ht]
    \centering
    \caption{Participants' Performance By SMT in Task 2.}
    \label{fig:participantObsTask2}
    
    \newcolumntype{R}{>{\RaggedLeft\arraybackslash}X}
    \newcolumntype{L}{>{\RaggedRight\arraybackslash}X}
    \rowcolors{2}{white}{gray!25}
    \begin{tabularx}{\columnwidth}{X R R R R R }
        \toprule
        \multicolumn{2}{r}{Completed} & SMT Used &  Secondary Source & W.A Method & Time\\
        \midrule 
        P12 &   $\checkmark$ &     HCP VS  &             &                &      4:37        \\
        P09 &   $\checkmark$ &     HCP VS  &   $\checkmark$ &   $\checkmark$ &   7:15 \\
        P03 &   $\checkmark$ &    HCP VS  &   $\checkmark$ &                 &  12:13 \\
        P16 &   $\checkmark$ &    HCP VS  &                &   $\checkmark$ &   16:44 \\
        P11 &   $\checkmark$ &    HCP VS  &   $\checkmark$ &   $\checkmark$ &   21:03 \\
        P21 &   $\checkmark$ &    HCP VS  &   $\checkmark$ &   $\checkmark$ &   28:44  \\
        P01 &                & HCP VS  &   $\checkmark$ &   $\checkmark$ &   $>$30:00 \\
        P10 &   $\checkmark$ &    Infisical   &                &                &   1:04         \\
        P20 &   $\checkmark$ &    Infisical   &                &                &   2:00            \\
        P05 &   $\checkmark$ &     Infisical  &                &                 &   4:04    \\
        P04 &   $\checkmark$ &     Infisical  &                &   $\checkmark$ &   5:18 \\
        P07 &   $\checkmark$ &    Infisical  &   $\checkmark$ &                &   16:23 \\
        P02 &                & Infisical  &   $\checkmark$ &   $\checkmark$ &   $>$30:00 \\
        P14 &                & Infisical  &                &   $\checkmark$ &   $>$30:00 \\
        P08 &   $\checkmark$ &     Doppler  &                &   $\checkmark$ &   7:50 \\
        P13 &   $\checkmark$ &    Doppler  &   $\checkmark$ &   $\checkmark$ &   13:20 \\
        P06 &   $\checkmark$ &    Doppler  &   $\checkmark$ &   $\checkmark$ &   13:28 \\
        P19 &   $\checkmark$ &    Doppler  &                &                &   22:51             \\
        P17 &   $\checkmark$ &    Doppler  &   $\checkmark$ &   $\checkmark$ &   23:10 \\
        P15 &   $\checkmark$ &    Doppler  &   $\checkmark$ &   $\checkmark$ &   23:23 \\
        P18 &   $\checkmark$ &    Doppler  &   $\checkmark$ &   $\checkmark$ &   29:05 \\
        \midrule
        Overall & 18  &   & 12     & 14     & 16:18$^\dagger$  \\ 
        \bottomrule
    \end{tabularx}
    \RaggedLeft
    $^\dagger$ Average
\end{table}
Our findings indicate that the number of participants that deviated from the expected path, as described in Section~\ref{sec:StatsAnalysisMethods}, doubled in count from Task 1 to Task 2. More than two-thirds of participants struggled to solely stick with the provided tool documentation and the expected methods for each task. We emphasize this as a focus in our findings since Task 1 requires several more individual steps than Task 2. Task 1 requires participants to remove \texttt{CompanySecretToken} from \texttt{SecureSecret.py}, store \\
\texttt{CompanySecretToken} in the SMT web UI \textit{CompanySecretStorage}, login and initialize into the SMT CLI, and finally print the value of \texttt{CompanySecretToken} from the command line. Task 2 on the other hand only requires the participant to input one command in the SMT CLI which performs the injection run process as shown in Figure~\ref{fig:hcpCLITask2}. It is unclear what specifically contributed to different paths participants took for each task by only looking at our data from Tables~\ref{fig:participantObsTask1} and ~\ref{fig:participantObsTask2}. Therefore, we use Sections~\ref{sec:WrapUp}, ~\ref{sec:ToolDocSpecificChallenges}, and  ~\ref{sec:SecSources} to find explanations for why participants experienced increases in measured difficulty and deviations from solely relying on tool documentation from the SMTs.

\subsubsection{Task Performance Comparing SMTs}
\label{sec:TaskStatsCompare}
Here, we briefly present differences in participant performance statistics we noticed between the three SMTs for each task.

\textbf{Task 1:} First, we examine how SMTs vary in their time-to-completion. Doppler (med. 11.9
minutes, $sigma=6.41$) and Infisical (med. 10.7 minutes, $sigma=6.84$) have task distributions with a center approximately double the median completion time of HCP
Vault Secrets (med. 5.58 minutes, $sigma=4.25$). None of the seven HCP Vault Secrets participants accessed secondary sources or tried workaround methods. As P16 noted, the tool was straightforward to use: \textquote[P16]{I liked how easy it was to create a token, to see the token \dots new developers are going to join and may use it more easily. There’s no complexity.} Whereas Infisical had the most people trying secondary sources and workaround methods. Because we only have seven participants in each group, statistical power is quite limited, and hypothesis tests should not be considered definitive. Nevertheless, we performed a Kruskal-Wallis test out of curiosity.
We found a statistically significant difference in distributions ($H=6.50$, $P=0.39$) with
a ``large'' effect size ($^2=0.250$).

\textbf{Task 2:} We repeat the same analysis for Task 2. Unlike Task 1, we had three participants who failed
to complete the task. Two of those participants used Infisical, which was the same SMT with the most participants not completing Task 2. For example, P14 struggled to understand how the tool linked the web portal to the local Python file, stating: \textquote[P14]{I wanted to link the Python file to the Infisical path where we store the key \dots I did not know just running the file would connect those two [Infisical web portal to the Python file].} Infisical also had the lowest number of participants who accessed secondary sources. We arbitrarily assigned a value of 30 minutes time for these
participants, though we find using a higher value did not change the statistical
conclusions. Doppler (med. 22.9 minutes, $sigma=7.53$), Infisical (med. 5.3 minutes,
$sigma=12.85$) and HCP Vault (med. 16.74 minutes, $sigma=9.95$) vary widely in
median but largely have overlapping distributions. We also performed a Kruskal-Wallis
test on these times, finding no statistically significant difference in the distributions
($H=1.41$, $P=0.495$).  Again, with low sample sizes these findings should be considered provisional.

\subsection{Tool Usability Feedback}
\label{sec:WrapUp}
Participants highlighted both positive aspects and suggestions for the SMTs from the study, indicating how different functionalities influenced their success in using the tool and completing the tasks. The copy to clipboard functionality was appreciated by P19, who mentioned they can easily copy to the clipboard and paste it in the \textit{``terminal directly''} along with the ease of creating tokens as mentioned by P16,
\textquote[P16]{I liked how easy it was to create a token, to see the token and it's like if there's a company that's using this application, new developers are going to join and may use it more easily. There’s no complexity.}
P21 also mentioned that the web dashboard was more helpful while working on the task, with its \textit{``UI is particularly helpful''} in managing secrets and allowing users to view all applications at once.
Many saw the SMT as a \textit{``good option for managing secrets''} with several others expressing interest in using it for future projects and exploring its integration capabilities. 
Participants also compared the tool to others, some felt it is a strong alternative to Git Lab secrets, while others mentioned preferring established cloud options like AWS for its documentation and familiarity like P15 stated, 
\textquote[P15]{I’m more of an AWS guy and it already has something like KMS. Like if you use AWS, KWS, EC2, you really don't have to do anything else, it is all in one place. I don’t think I’d want to use Doppler. Also, the documentation for KMS is really nice, people have been using it for a while.}

In terms of personal versus professional usage, participants like P01 generally felt the tool was more suited for \textit{``company or team-based projects rather than personal use''}.

P09 mentioned that the \textit{``pricing structure of the tool''} is one of the deciding factors for adoption. Some participants like P04 also mentioned that the tool was easy to set up and had clear documentation and an intuitive user interface. P13 felt that the dashboard and command-line interface further improved the overall usability by stating, 
\textquote[P13]{I think the tool was very helpful in managing the secrets. I like the way we can interact with the dashboard like they have created a dashboard that then we can interact via the visual interface and we can import the secrets as well if we don't want to use the command line. Whereas people who love the command line, have a dedicated command line as well. So I like that idea pretty much.}

Participants also pointed to various features that supported their needs where P16 appreciated the \textit{``availability of API, CLI and customization''}. Whereas, P08 also pointed out that for projects, relying on a tool like Doppler \textit{``could create dependency and difficulties in migration''}. P01 faced login issues, mentioning they \textit{``didn't know how the login works''}. P15 also highlighted problems with the save feature, stating they \textit{``didn't really like it''} and P04 also pointed out issues with handling special characters, stating it \textit{``could improve on string escape characters''}.

The study also revealed usability differences between SMTs in Section ~\ref{sec:TaskStatsCompare}, where HCP Vault Secrets had faster completion times and fewer instances of participants seeking secondary sources compared to Infisical. This could indicate that its workflow is easier or more intuitive for users. In contrast, Infisical had a higher rate of task failure and more users relying on secondary sources, which might point to usability challenges like input handling limitations, and flexibility. The flexibility offered by different user interfaces could play a role in adoption, though further research is needed to better understand how these factors impact long-term usage and success.

\subsection{Tool Documentation Challenges}
\label{sec:ToolDocSpecificChallenges}
In this section, we specifically focus on challenges and experiences directly reported from using SMT tool documentation and web UI to complete each task. 

\subsubsection{Documentation Content}
\label{sec:ToolDocSpecificContent}

Tool documentation was widely reported as providing enough information and being generally helpful by participants during Task 1. 13 participants even cited specific documentation they found helpful. For example, both P04 and P10 mentioned that the Infisical CLI quickstart guide was helpful for managing secrets and gave basic steps to do so. P03 reported that HCP Vault Secrets was \textit{``easy to navigate''} and appreciated the usage of images in documentation. P18 and P19 appreciated CLI command descriptions in Doppler's CLI Guide. 

There were a few instances when participants mentioned a lack of clarity over CLI descriptions and usage during Task 1. P09, P11, and P19 expressed confusion on applying CLI secrets commands when command line arguments were not explicitly defined on the same page in the documentation. Participants reported confusion over CLI usage when they felt the content did not sufficiently cover the relationship between CLI commands and secret management processes. Both P04 and P19 felt the content they used did not fully address immediate actions required to complete the initialization and login process for using the CLI.

Participants who attempted workaround methods for Task 1 mostly tried CLI commands to export secrets and also the injection run command which was needed for Task 2. None of the participants were aware they were going to inject secrets before Task 2, so some of them used the injection run command to perform injection to complete Task 1. P15 mentioned the following as to why they used the injection run command for Task 1: \textquote[P15]{So my task was basically to output the secret, right? So the first thing that I did was like doppler run\dots}

For Task 2, participants largely reported tool documentation as being much less helpful and consequently reporting Task 2 as more challenging than Task 1. 17 participants reported that content for the injection run command contributed to challenges they experienced when completing Task 2. Participants mentioned the following problems: lack of sufficient information for relevant command line arguments, lack of a relevant set of use cases in examples provided, and inconsistent documentation that covers secret injection.

\textbf{Command Line Arguments:} Several participants mentioned that command line arguments were introduced in the tool documentation for the injection run command. However, they were confused reading because they felt there was a lack of explicit information on how to apply them and how they related to the injection run command. P09 and P16 both referred to injection run command documentation from HCP Vault Secrets CLI page. They did not understand the \texttt{duration} flag shown at the end of the command in the documentation.
P16 could not figure out how to make the \texttt{duration} flag work with the injection run command. P16 also mentioned a lack of understanding towards another flag used in the documentation:
\textquote[P16]{And the first thing, if you see, they have mentioned app, my-app, the application you want to pull all secrets from. But you don't understand its purpose or it doesn't say what happens if you don't do it. I had to try it myself.} 

\textbf{Example Use Cases:} Six participants explicitly stated examples provided from the tool documentation were a challenge when figuring out how to apply the injection run command to the study Python file, ~\textit{SecureSecret.py}. Before our full study, injection run command examples shown in all of the SMTs tool documentation did not explicitly present a Python example, unlike other languages and frameworks such as Flask, Golang, NPM, and command line functions like echo. Participants wanted more clarification from tool documentation on how languages or frameworks used in the examples are simply just examples and not required to run the command itself. P17 explained how they were initially confused with the explicit injection content in Doppler's Secret Access CLI Guide since it primarily used an example with Node Package Manager (NPM) without any other examples present:
\textquote[P17]{This npm start is just a command example\dots They should have mentioned that npm start is just an example you can type in your command instead of start. That got me a little confused. Otherwise, the task was easier but it took me some time to figure out the exact solution.}

\textbf{Inconsistent Command Presentations:} Several participants mentioned they observed discrepancies in relevant command presentations used in content describing secret injection across different documentation pages. Multiple participants that used either Inifisical or HCP Vault secrets reported that both tool's quick start guides and their CLI documentation pages differed in both the examples and arguments present for their injection run command descriptions. P05 mentioned that while Infisical's injection run command page provides a bare example without flags or arguments, the Infisical quickstart guide only provides examples with sample flags and arguments. P05 later explained that led them to try the injection command with incorrect flags and arguments because they did not know beforehand those were not necessary, thus received errors at first. Participants that used HCP Vault Secrets faced this challenge even with the existence of added dashboard documentation that provides a specific Python3 example for the injection run command, as shown in Figure~\ref{fig:hcpWebPortalPython}. For participants who noticed the added dashboard content, they mentioned that the inclusion of the Python injection run command example helped them complete the task. Specifically, participants P03 and P09 found the HCP Vault Secrets Python3 example helpful for understanding how to use secret injection feature. Participant P03 noted, \textquote[P03]{One is like not having to type out the credentials again and again, and not having to hard code the credentials which is not ideal. Scoring it in the environment variables with secret injection is a lot safer.} Similarly, P09 explained, \textquote[P09]{They have this feature and I can inject my secret. So it is like completely decoupled from my main application.}

However, they clarified it was not intuitive to expect secret injection documentation within the web dashboard or outside of the public tool documentation. P11 noticed the Python example in the HCP Vault Secrets dashboard and stated that \textit{``maintaining two different pages and they have two different contents, obviously confuses people''}. P03 echoed a similar response:
\textquote[P03]{I would say the first one was easier because the links to the documentation were available\dots In my mind, I was assuming that there will be similar documentation available somewhere. And it did not strike me that I can just take a look at the portal.}

We believe insufficient information, a lack of relevant use cases in examples, and inconsistent documentation contributed to participants deviating from strictly following tool documentation. From our interview responses, participants reported struggling with understanding how to perform secret injection in Python when reading tool documentation command line arguments and examples. Inconsistent presentations of injection run commands also confused participants as to how to perform secret injection in Python.

\subsubsection{Documentation Structure and Formatting}
Here we report responses from participants towards the structure and formatting of the tool documentation. We also briefly report specific navigation difficulties participants experienced with the search functionality in the tool documentation. 

\textbf{Structure:} The structure of the tool documentation received generally positive feedback. Participants appreciated the overall clarity in structure, including well-defined headers and organized sections like P18 mentioned, \textit{``everything is labeled nicely''}. Both P04 and P20 felt the web UI for the tool documentation was clear and minimalist in design. P04 specifically appreciated when titles, headers, and tabs were used to separate topics. 

\textbf{Content Format Preferences:} Regarding content format, participants had mixed preferences. Some participants preferred text for its searchability and ease of reference. When asked the reason for preferring text, P09 explained they \textit{``prefer reading text over videos''} as it allows them to search for keywords and scan information faster. Other participants found code examples and videos beneficial for visual learning. For coding issues, P21 mentioned, \textit{``text with code blocks is very helpful''}. P17 explained their preference for videos:
\textquote[P17]{First reference would be video because if I want to get help for something then it would be better for me to see that the other person is doing the same thing and explaining it in a video.}

Few participants mentioned their content preference depended on the complexity of the task at hand. P01 stated, \textit{``minor problem may be a text, a bigger problem somewhere in between text and video''}. Overall, participants held mixed preferences regarding format, suggesting that offering mixed formats could better support varied user needs.

\textbf{Navigation Difficulties:} During both tasks, participants reported specific navigation difficulties for the tool documentation. P04 felt that the \textit{``search functionality could be improved''} in the documentation, as they struggled to find specific terms or commands, and mentioned that the \textit{``ease of search was missing.''} Participants found it most challenging to use the search functionality from the tool documentation to find sufficient information for secret injection, as P04 mentioned:
\textquote[P04]{\dots when I typed injection, it did not give me the desired response. So I think it is good that they are returning an LLM response, but I would also appreciate if they would have returned a link to where I could find what I actually wanted. They're basically giving me a ChatGPT kind of response, but I would also appreciate the source of that thing.}

As navigation difficulties increased for participants during Task 2, their time and effort also increased as they needed to perform more troubleshooting and exploration.

\subsection{Secondary Sources}
\label{sec:SecSources}
Here, we focus on participants responses towards why they accessed secondary sources and how helpful they reportedly found secondary sources. We do not analyze the secondary sources in-depth themselves (e.g., content, structure, formatting, or source of origin). We only provide some examples and briefly discuss relevant content from examples participants mentioned where relevant.

Only two participants during Task 1 accessed secondary sources online for further assistance, as shown in Table~\ref{fig:participantObsTask1}. One of those participants, P14, mentioned they accessed a secondary source to reduce confusion for the injection run command by finding more examples. As we previously mentioned in Section~\ref{sec:ToolDocSpecificContent}, a few participants from Task 1 attempted the injection run command intended for Task 2 because they thought that command would produce the intended output required for Task 1. This lead P14 to seek secondary sources on how to use the injection run command during Task 1. We explain implications behind how this can lead to complications broadly for users adopting new software technology through tool documentation in Section~\ref{sec:DiscNegativeExperiences}.

Seeking further assistance on how to perform secret injection in Python became the common reason for participants accessing secondary sources during Task 2. As shown in Table~\ref{fig:participantObsTask2}, twelve participants searched for and accessed secondary sources outside the tool documentation during Task 2. Ten of those participants explained they needed information which was not addressed in the tool documentation. P06 stated the tool documentation was not necessarily confusing, but they needed \textit{``maybe a little more description''}. P21 felt the examples provided were not elaborate for completing Task 2. This reasoning led participants to search the web for secondary perspectives on how to inject secrets with their SMT. P18 described this process: \textquote[P18]{I scrolled quickly through the main documentation and I was stuck. So it's like whenever you are stuck at something, you need some other perspective.}

Finding solutions from others online motivated participants in our study to use sites like StackOverflow. P15 felt that while the Doppler CLI Guide was \textit{``really good to get started''}, StackOverflow is better for finding the \textit{``specific piece of code you are actually looking for''}. P21 mentioned they appreciate added background and debugging advice from users on StackOverflow. Participants who searched for secondary sources often indicated those sources helped them more than the tool documentation when they received that added bit of information or example usage which they felt was missing from the tool documentation, specifically for Task 2. P06 googled broadly for information for running Doppler with Python files and mentioned that a Dev Community blog post~\cite{DopplerPythonDevBlog} gave better descriptions for running the injection run command with Doppler for Python files.

However, more often than not, participants would either try multiple search queries and multiple distinct secondary sources before either finding one that helped or eventually  going back to the tool documentation for Task 2. P03 even mentioned that when they did try to Google for secondary sources, the results returned \textit{``were totally irrelevant.''} P15, while searching for sources, clicked on a blog post by Medium Corp.~\cite{DopplerInjectionKubernetesSecondary} which they believed would give relevant advice for secret injection. However, P15 later clarified that the blog \textit{``was very complex and covered kubernetes. It didn’t actually tell me what I wanted.''}

As mentioned in Section~\ref{sec:TaskSelection}, Doppler provides a hyperlink within the Development/Editors Section~\cite{DopplerEditorsPython} for Python which directs users to a external doppler-env-package page~\cite{DopplerPython} that shows an injection run command example in Python. While three of the Doppler participants specifically mentioned that the Python example for secret injection from the doppler-env-package page helped, finding the page was a challenge. P15 mentioned that doppler-env-package page~\cite{DopplerPython} \textit{``wasn’t really anywhere in the documentation.''} P18 on the other hand noticed the hyperlink for the doppler-env-package page, however the surrounding headers and descriptions mentioned configuration and installation steps. Therefore, P18 early only thought any hyperlinks within that section were \textit{``only about installation.''} They did not reach the doppler-env-package page until Googling for more documentation and being directed to it. 

The motivation from participants to acquire secondary sources increased noticeably when the tool documentation provided by the SMTs were perceived to not be sufficient enough in their explanations or examples, or provided inconsistent information. Even with the direct Python examples of injection run commands from HCP Vault Secrets web dashboard and Doppler's external doppler-env-package page, participants experienced difficulty in finding that information. HCP Vault Secrets participants did not expect relevant content for their tasks to be in their web dashboard, and Doppler participants could not find the doppler-env-package hyperlink which is mentioned in a section not specifically related to secret management. Other secondary sources were only helpful if they gave descriptions or code snippets participants felt that the tool documentation did not provide. However, most of the secondary sources accessed by participants were not much more helpful than the tool documentation, leading participants back to the tool documentation to finish Task 2.

\section{Discussion}
We discuss how our findings provide context for why new users may report challenging experiences when first setting up and using SMTs to manage secrets. We discuss SMT-specific conclusions towards new user SMT experiences in Section~\ref{sec:DiscRQ1}. Then, we discuss broadly applicable conclusions towards tool usability and documentation in Section~\ref{sec:DiscRQ2}. We use both our participant performance and interview findings to complement each other. We conclude with providing recommendations for SMT providers when designing SMT tool documentation and interfaces in Section~\ref{sec:DiscFutureWorkRec}.

\subsection{RQ1: New Users and SMT Usage}
\label{sec:DiscRQ1}
Participants joined our study with prior experience in secret management and related programming concepts from our eligibility criteria. Some held experience using SMTs not from this study while others used a SMT for the first time during this study. We relieved participants of typical onboarding duties and provided a minimal example use case to remediate a hard-coded secret as described in Section~\ref{sec:StudyDesign}. However, we did not not observe noticeable differences in participant task completion time and their ability to complete tasks with using workaround methods or secondary sources. Therefore, our work serves as a baseline for the best-case scenario for how new users would realistically perform with the set up and usage of SMTs to centrally secure their secrets.

Our findings support prior work towards identifying how developer experiences using new tools can be drastically impacted by the quality of available resources present~\cite{gorski2018developers,basak2022practices,krause2023pushed,acar2016you}. Even in a laboratory study of the best-case scenario, the ability of participants to effectively and securely remediate a hard-coded secret changed drastically between each task. The number of new users that accessed secondary sources or attempted workaround methods doubled from Taks 1 to Task 2. Even though participants completed Task 1 with relative ease, Task 2 is the process which helps ensure users can securely run local applications without hardcoding secrets. Both tasks were designed to represent fundamental processes for SMTs which aim to reduce security vulnerabilities for managing secrets, while being short enough to complete consecutively with four CLI command prompts as shown in Figures~\ref{fig:hcpCLITask1} and ~\ref{fig:hcpCLITask2}. Yet, even in our laboratory study with prior secret management experience, participants spent approximately half an hour and deviated from official tool documentation guidance just to complete both tasks. Even if participants gained incidental knowledge to complete Task 2 during Task 1, Task 2 is designed so that participants still needed to perform secret removal and secret storage steps with the SMT in Task 1. Therefore, our conclusions do not change when observing the total experience for both tasks. If participants in our laboratory setting struggled to effectively perform both tasks, then it is likely new users in a real world setting experience similar challenges with even less guidance on how to effectively use SMTs to securely manage secrets, which demonstrates a contribution to adoption burdens reported by developers in prior work~\cite{krause2023pushed,basak2023challenges}.

\subsection{RQ2: Functionality and Documentation Factors that Impact SMT Usage}
~\label{sec:DiscRQ2}
We examine the functional, positive, and negative aspects of participants' experiences using SMTs and tool documentation to manage secrets. Participants reported that well-structured documentation supported task completion, while unclear CLI instructions and limited examples created challenges.
\subsubsection{Functionality Impact}
We performed statistical analyses to see if we could identify any differences in participant performances across the three SMTs. While we do not claim definitive statistical significance, we found median completion times across SMTs varied widely, but largely overlapped in range. For example, in Tables~\ref{fig:participantObsTask1} and ~\ref{fig:participantObsTask2}, three and two participants completed Task 1 and 2 in under four minutes, respectively. On the flip side, three and nine participants took at least 20 minutes to complete Task 1 and 2, respectively. Thus, we see the possibility for participants to effectively use SMTs to secure secrets quickly, but we also see many other participants look at the same documentation and take considerably longer. Therefore, we reason the SMTs in our study were functionally indistinguishable compared to each other when participants used them to manage secrets. However, by functionally indistinguishable, we refer specifically to their usability features, such as those highlighted in Section~\ref{sec:WrapUp}, and not functional equivalence between the SMTs. Differences in participant task completion times likely reflect SMT usability variations between each task. While we acknowledge that our sample size limits statistical significance, our work serves as an appropriate exploratory study addressing first-time user experiences with SMTs~\cite{basak2023challenges,krause2023pushed}.

Participants pointed out several UI and interface features that influenced their experience with SMTs. They appreciated intuitive web interfaces, copy-to-clipboard options for command snippets, and well-structured layouts with clear headers, all of which made their tasks easier. These usability improvements align with findings from recent studies, which highlight the importance of UI design and guided workflows in secret management tools~\cite{somasundaram2024unified}. We also noticed that those who found the UI intuitive had fewer difficulties managing secrets, while those who struggled with the interface took longer to complete tasks and felt more frustrated. Previous work has similarly shown that poorly designed interfaces and unclear documentation can cause secrets to be mismanaged or accidentally leaked~\cite{huang2024your}.  
This supports our findings that while SMTs may offer similar functionality, the quality of SMT tool UI design has a significant impact on overall usability and adoption by helping users navigate and complete tasks more smoothly~\cite{basak2023secretbench}.

\subsubsection{Documentation Positive Factors and Experiences}
Participants provided multiple positive statements about the 
tool documentation they used with the SMT, mostly while performing Task 1. Similar to prior efforts~\cite{acar2017comparing,smith2020can,indela2016helping,nassif2021generating,lethbridge2003software}, participants in our study largely appreciated tool documentation that simplified their process of finding relevant information and provided just enough information to complete each task. 
Additionally, several participants also expressed a preference for video-based tutorials, indicating that these visual formats are effective and make it easier to understand and follow the instructions, including for non-expert users of a new tool~\cite{liu2024having}.
Participants felt that the tool documentation provided enough information to complete Task 1.  
The inclusion of quick starter guides, and structural features like well-defined headers and organized sections were common aspects participants felt made their experience easier. We provide further insight from prior work in directly observing that as participants felt more positive about tool documentation, they completed their task in fewer time with less challenges along the way. Most importantly, we saw that participants were less likely to deviate the expected path while using SMTs to manage secrets when they reported positively about tool documentation.
Thus, decreasing the risks of traversing through irrelevant information, implementing insecure methods, or an inability entirely to learn how to use new tools.

\subsubsection{Documentation Negative Factors and Experiences}
\label{sec:DiscNegativeExperiences} 
Participant feedback revealed that documentation challenges were significant barriers, leading us to shift our focus toward examining the usability of SMT tool documentation. Our findings support related efforts by continuing to highlight tool documentation challenges such as insufficient coding examples~\cite{aghajani2020software,aghajani2019software,nassif2022casdoc,acar2017developers,acar2017comparing,subramanian2014live}, incorrect or ambiguous explanations~\cite{uddin2015api,wen2019large,middleton2020data,chen2009empirical,acar2017developers,sohan2017study,treude2016augmenting} and the growing use of LLMs to generate documentation, which may introduce inaccuracies and further mislead users~\cite{bhat2024llms,hsieh2023tool}.
However, our work extends prior efforts by exploring challenges participants report with using SMT-specific tool documentation covering CLI commands. New users already face a learning curve when first adopting SMTs~\cite{krause2023pushed,basak2023challenges}. If new users encounter tool documentation with insufficient descriptions for relevant CLI command line arguments, or without a broad range of example use cases for CLI commands, they are less likely to know how to effectively use the SMT CLI functionality. Therefore, new users are also less likely to understand how to securely manage their secrets with SMTs when reading provided tool documentation. We discuss recommendations that SMT providers should consider when writing tool documentation to reduce participant confusion when learning SMT CLI commands in Section~\ref{sec:DiscFutureWorkRec}.

Our work also provides more context towards the experiences new users have when searching secondary sources for technical assistance as highlighted in prior work~\cite{arya2023programmers,baltes2020contextual,robinson2022error,storey2024disruptive,acar2016you,parnin2012crowd}. Most relatedly, Acar et al.~\cite{acar2016you} observed that Android code written by developers produced different levels of security and functionality depending on the information source the developer used for assistance. In our study, participants complained that information from secondary sources was either no different than what the tool documentation provided, overly complex for their task, or completely irrelevant. In a real world setting, new users learning a tool search for information sources and need to determine if the source they are looking at is proposing methods that will help them achieve the expected outcome they require. The extent of harm for applying incorrect, or worse insecure, methods increases if new users feel inclined to spend more time exploring secondary sources after unsuccessfully trying official tool documentation. Therefore, making it more difficult for new users to effectively use or adopt new tools. We believe our findings for participants using SMT tool documentation add context to adoption burdens reported by developers~\cite{basak2023challenges,krause2023pushed}. Specifically, if SMT tool documentation is challenging to use, and secondary sources discussing SMTs only add more confusion, then developers may be less willing to further adopt SMTs. Developers interested in adopting SMTs may have challenges using SMT tool documentation to manage their secrets or navigating different information sources to find relevant solutions for their use cases.

\subsection{Recommendations}\label{sec:DiscFutureWorkRec}
Here, we offer recommendations to SMT providers regarding  tool functionality and documentation. 

\textbf{Dedicated Sections for Technical Terms:} Many participants recommended including dedicated sections in the documentation to explain technical jargon, such as command flags, subcommands, and arguments. These sections could provide clear definitions, examples, and a glossary to help users, especially beginners, navigate the documentation more effectively. 

\textbf{Visual Aspects for Demonstration:} Additionally, participants also suggested including visual aspects in the documentation like flow diagrams to demonstrate how commands interact with values, arguments, and output. These diagrams could provide a clear logical flow for the key tasks in the SMTs like logging in, initializing processes, and accessing secret projects. While resolving issues, users could also benefit from having debug examples, code snippets, and guidance. 

\textbf{Structural Improvements:} Another recommendation is regarding the structuring of CLI documentation with a base example at the top, followed by examples with subcommands, flags, and arguments. This approach keeps all relevant information on one page, reducing the need for users to search elsewhere, and improving the overall user experience by minimizing lookup time. An effective structure and formatting with clear headings, a logical flow, and tables to represent a list of commands, subcommands, and argument explanations would help users quickly access and understand the information. Also, adding hyperlinks or resources in the documentation to guide users on installing necessary programming environments or dependencies would help beginners to quickly set up and follow the rest of the instructions to complete their tasks. 

\textbf{Clear Placeholders in Commands:} Participants reported confusion over which CLI argument parameters were necessary to replace or use verbatim. Documentation authors should use obvious placeholders (e.g., \texttt{{your\_project\_name}})  to help users avoid confusion. 

\textbf{User-friendly Features:} Providers are also recommended to continue to include user friendly features like copy to clipboard for ease of use and also consider adding search functionality and quick navigation links to improve the tool usability and help users complete their tasks more efficiently. 

\textbf{Incorporating Visual Media:} Many participants expressed a preference for video-based tutorials as part of the documentation, as this format can help them better understand how to use tool features. Therefore, providers should aim to incorporate visual elements such as interactive demos and walkthroughs that demonstrate command usage, argument flows, common troubleshooting steps, and mirror real usage scenarios.

\textbf{LLM Integration:} Participants struggled when documentation lacked concrete examples or when search features failed to return relevant information. While LLMs can help generate diverse, contextual examples, participants also noted confusion when responses were vague or lacked sources. We recommend that SMT providers use LLMs to augment, not replace, documentation by generating accurate, targeted examples that are clearly linked to official sources.

\section{Conclusion}
In our exploratory qualitative two-step laboratory study into the usage of SMTs, we observed how even new users with prior secret management experience and an easier onboarding experience struggled to perform secret management tasks using a SMT. We identified specific challenges relating to tool documentation content, structure, and formatting that participants reported impacted their ability to effectively and securely remediate a hard-coded secret. Participants responded to tool documentation and interface challenges by deviating towards accessing secondary sources and attempting workaround methods, while cycling back to tool documentation and experiencing added difficulty. We make recommendations for SMT vendors to improve the ecosystem of guidance for new SMT users.

\section*{Acknowledgments}
This material is based upon work supported by the
National Science Foundation under Award Nos. CNS-2055554 and CNS-2206865. 
We want to thank Anna Lena Rotthaler for her valuable feedback and interest in this research. We are thankful to all study participants for generously sharing their time and insights. We also appreciate the anonymous reviewers for their constructive suggestions and guidance.

\bibliographystyle{ACM-Reference-Format}
\bibliography{bibliography}

\appendix

\section{Interview Questions}
\label{sec:interview questions}
\footnotesize
\noindent\textbf{Warm-up Questions: Prior experience with managing secrets}
\begin{enumerate}
    \item Do you have any prior experience with managing secrets? (E.g., managing secrets for an industry role? Academic project?)
    \item What type of secrets did you manage?
    \item Did you use any specific tools or perform specific management practices?
    \end{enumerate}
\textbf{Task 1: Overall Experience}
\begin{enumerate}
    \item How would you describe your overall experience of performing this task?
    \item Is there anything that you did not like or find difficult for this task?\\
    a) Vice Versa?
\end{enumerate}
\textbf{Task 1: Negative Emotions}
\begin{enumerate}
    \item Were there any parts during this task exercise where you experienced any negative emotions? \textit{(This can include annoyance, confusion, fatigue, frustration, or any form of disapproval/dissatisfaction with either the tool, the advice you looked at, or the task itself.)} \\
    a) If so, please explain why you felt those emotions.
\end{enumerate}
\textbf{Task 1: Tool Documentation}
\begin{enumerate}
    \item Did you experience any difficulties or challenges when using code snippets, examples, or explanatory text from the tool documentation as resources to complete the task? \\
    a) If so, why? What made using these specific qualities of the content challenging?
    \item Did you experience difficulty in mentally connecting the process for what you were assigned to do in the task with how the respective content is provided in the tool documentation? \textit{(To rephrase: Was it challenging to understand how to complete the task assigned to you as you were reading the content provided in the tool documentation?)} \\
    a) If so, please explain
    \item Did you experience any challenges in finding the information that allowed you to complete or progress through the task? \\
    a) If so, please explain what challenges you faced in finding the information.
    \item Did the structure of the tool documentation content (e.g., section headers,  usage of formatting like bullet points or lists, toolbar format, etc) present any challenges for you in finding information and completing the task? \\
    a) If so, please explain. and b) Vice Versa?
    \item Did the format of the advice content (as it relates to text, pictures, videos) present any challenges to you or make it harder to complete the task? \\
    a) If so, please explain. \\
    b) Vice Versa?
\end{enumerate}
\textbf{Task 1: Secondary Sources}
\begin{enumerate}
    \item If you looked up advice from sources external to the tool documentation, why did you search for other resources? \\
    a) What specifically did you look for in secondary sources that were not present and/or made clear within the help documentation?
    \item Were there any external sources that you preferred or found more helpful towards completing the task than the tool documentation?\\ a) If so, what were these sources and why did you find them more helpful?
\end{enumerate}
\textbf{Task 2: Overall Experience}
\begin{enumerate}
    \item How would you describe your overall experience of performing this task?
    \item Is there anything that you did not like or find difficult for this task? \\
    a) Vice Versa?
\end{enumerate}
\textbf{Task 2: Negative Emotions}
\begin{enumerate}
    \item Were there any parts during this task exercise where you experienced any negative emotions? \textit{(This can include annoyance, confusion, fatigue, frustration, or any form of disapproval/dissatisfaction with either the tool, the advice you looked at, or the task itself.)} \\
    a) If so, please explain why you felt those emotions.
\end{enumerate}
\textbf{Task 2: Tool Documentation}
\begin{enumerate}
    \item Did you experience any difficulties or challenges when using code snippets, examples, or explanatory text from the tool documentation as resources to complete the task? \\
    a) If so, why? What made using these specific qualities of the content challenging?
    \item Did you experience difficulty in mentally connecting the process for what you were assigned to do in the task with how the respective content is provided in the tool documentation? \textit{(To rephrase: Was it challenging to understand how to complete the task assigned to you as you were reading the content provided in the tool documentation?)} \\
    a) If so, please explain
    \item Did you experience any challenges in finding the information that allowed you to complete or progress through the task? \\
    a) If so, please explain what challenges you faced in finding the information.
    \item Did the structure of the tool documentation content (e.g., section headers,  usage of formatting like bullet points or lists, toolbar format, etc) present any challenges for you in finding information and completing the task? \\
    a) If so, please explain. and b) Vice Versa?
    \item Did the format of the advice content (as it relates to text, pictures, videos) present any challenges to you or make it harder to complete the task? \\
    a) If so, please explain. and b) Vice Versa?
\end{enumerate}
\textbf{Task 2: Secondary Sources}
\begin{enumerate}
    \item If you looked up advice from sources external to the tool documentation, why did you search for other resources? \\
    a) What specifically did you look for in secondary sources that were not present and/or made clear within the help documentation?
    \item Were there any external sources that you preferred or found more helpful towards completing the task than the tool documentation?\\ a) If so, what were these sources and why did you find them more helpful?
\end{enumerate}
\textbf{Wrap-up Questions}
\begin{enumerate}
    \item Could you see yourself using this SMT for managing secrets within your own workflow? Or for your own projects? \\
    a) If not, can you please explain why not? and b) If yes, please explain, specifically, what features or aspects help with the adoption.
    \item Do you have any preferences for the types of sources you consult when needing assistance for technical tasks like the ones today? (including online or offline)
    \item Do you have any preferences for how you typically like to view advice from help resources? (e.g., just text, media included, etc)
    \item Do you have any last comments or questions about SMTs, tool documentation, or the study in general?
\end{enumerate}

\section{Participant Demographics}
\label{demographics}
\begin{table}[htbp]
    \centering
    \caption{Participant Demographics and SMT Assignment.}    \label{fig:participantExp}
    \begin{tabular}
    { p{1.1cm} p{1.3cm} p{1.5cm} p{1cm} p{1.5cm} }
        %\hline
        %\multicolumn{4}{|c|}{Participant Prior Experience} \\
        \toprule
        Participant No. & Professional Exp. & Academic Exp. & Gender & SMT Assignment\\
        \midrule
        P01 & Internship & Coursework & Man & HCP Vault \\
        P02 & N/A & Coursework & Woman & Infisical \\
        P03 & Internship & Coursework & Man & HCP Vault \\
        P04 & Full-Time & N/A & Man & Infisical\\
        P05 & Internship & Coursework & Man & Infisical\\
        P06 & Full-Time & Coursework & Man & Doppler\\
        P07 & Full-Time & Coursework & Woman & Infisical\\
        P08 & Full-Time & N/A & Man & Doppler\\
        P09 & Full-Time & N/A & Man & HCP Vault\\
        P10 & Internship & Coursework & Man & Infisical\\
        P11 & Internship & Coursework & Man & HCP Vault\\
        P12 & Full Time & N/A & Man & HCP Vault\\
        P13 & Full Time & Coursework & Man & Doppler\\
        P14 & Internship & Coursework & Woman & Infisical\\
        P15 & Full-Time & N/A & Man & Doppler\\
        P16 & Full-Time & N/A & Man & HCP Vault\\
        P17 & Full-Time & N/A & Man & Doppler\\
        P18 & Internship & N/A & Woman & Doppler\\
        P19 & Full-Time & N/A & Man & Doppler\\
        P20 & Full-Time & N/A & Woman & Infisical\\
        P21 & Full-Time & N/A & Man & HCP Vault\\
        \bottomrule
    \end{tabular}
\end{table}

\newpage

\section{Interview Codebook}\label{sec:InterviewCodebook}
\begin{table}[h]
\centering
\caption{Interview Codebook of Participants' Experiences and Wrap-up.}
\label{Interview data codebook2}
\footnotesize
\rowcolors{1}{white}{gray!10}
\begin{tabular}{p{2.5cm} p{5cm}}
  \toprule
  \textbf{Codes} & \textbf{Definitions} \\ \midrule
  Prior Orgs. and Roles & Types of orgs(e.g.,industry, academic, govt) and roles (e.g., full-time, intern) related to managing secrets (e.g., tokens, backend APIs)\\
  Secrets Managed & Types of secrets managed\\
  Tools and Practices & Specific tools or practices used\\
  Tool Usability Feedback & Reasons about using the tool in future and why/why not?\\
  General Source Lookup & Preferred sources for general technical problems\\
  Advice Format Preference & Preferred format for online advice (e.g., just text, media included, etc)\\
  Last Comments & Any last comments or questions about the study\\
  Security Implications & Concerns, comments, or observations about security of tool or task(security implications)\\
  Recommendations & Voicings of recommendation or suggestion about tool or doc helped to finish task\\
  General Tool Comments & Feedback on tool features (e.g., save button, UI design, copy/paste functionality)\\
  Chat GPT/AI Comments & Responses mentioning usage of ChatGPT\\
  Google Docs Advice Comments & Comments on using provided Google docs links\\
  Web Portal or Dashboard Advice & Comments on web portal/dashboard (tasks or sections)\\
  \bottomrule
\end{tabular}
\end{table}

\begin{table}[b]
\centering
\caption{Interview Codebook of Tasks 1 and 2.}
\label{Interview data codebook1}
\footnotesize
\rowcolors{1}{white}{gray!10}
\begin{tabular}{p{2.5cm} p{5cm}}
  \toprule
  \textbf{Codes} & \textbf{Definitions} \\ \midrule
  Overall Experience & Overall experience to complete the task\\
  Subcode: Overall & Captures general responses without specific positive or negative details\\
  Subcode: Negatives & Specific difficulties, challenges, or dislikes related to the task\\
  Subcode: Positives & Aspects participant liked, found easy about the task, documentation\\
  Negative Emotions & Expressions of negative emotions (e.g., confusion, annoyance, frustration, fatigue, dissatisfaction) noted directly or during responses\\
  Documentation Content Challenges & Issues with content such as code snippets, commands, UI examples, explanatory text, images, or videos\\
  Subcode: Challenges & Issues or dislikes participants encountered when using content from the main help documentation to complete the task\\
  Subcode: Helpfulness & Reasons the content was helpful or aspects the participant liked\\
  Task and Documentation Mental Connection & Difficulty of linking the task description and goal with the help documentation\\
  Subcode: Yes & If a participant says yes, and the reasoning why\\
  Subcode: No & If a participant says no, and the reasoning why\\
  Structure and Formatting Challenges & Issues or dislikes related to the structure or formatting of the main help documentation\\
  Subcode: Finding Helpful Advice & Challenges of participant broadly for finding advice\\
  Subcode: Structure & Challenges with the structure, such as page format, section headers, bullet points, tabs, or toolbars, encountered while completing the task\\
  Subcode: Format & Challenges with the format, including text, images, videos, or other media, encountered while completing the task\\
  Subcode: Helpfulness & Positive aspects of the structure, format, or advice that helped the participant complete the task\\
  Secondary Source Lookup & Mentions of secondary sources outside the materials provided in the study\\
  Subcode: Lookup & Reasons for accessing secondary sources and what the participant was seeking\\
  Subcode: Not Helpful & Reasons why the participant found secondary sources unhelpful or challenges encountered\\
  Subcode: Helpfulness & Responses for if the participant found the secondary sources more helpful and why/why not?\\
  \bottomrule
\end{tabular}
\end{table}

\end{document}